\documentclass[sigconf]{acmart}

\usepackage{algorithm}
\usepackage{algorithmic}

\setcopyright{none}
\renewcommand\footnotetextcopyrightpermission[1]{}
\begin{document}

\title{ ETF Portfolio Construction via Neural Network trained on Financial Statement Data}

\author{Jinho Lee}
\authornote{Corresponding Author.}
\affiliation{%
  \institution{Shinhan-AI}
  \city{Seoul}
  \country{Korea}
}
\email{jinholee@shinhan.com}

\author{Sungwoo Park}
\affiliation{%
  \institution{Shinhan-AI}
  \city{Seoul}
  \country{Korea}
}
\email{sungwoopark@shinhan.com}

\author{Jungyu Ahn}
\affiliation{%
  \institution{Shinhan-AI}
  \city{Seoul}
  \country{Korea}
}
\email{jungyahn@shinhan.com}

\author{Jonghun Kwak}
\affiliation{%
  \institution{Shinhan-AI}
  \city{Seoul}
  \country{Korea}
}
\email{jkwak@shinhan.com}

\begin{abstract}
Recently, the application of advanced machine learning methods for asset management has become one of the most intriguing topics. Unfortunately, the application of these methods, such as deep neural networks, is difficult due to the data shortage problem. To address this issue, we propose a novel approach using neural networks to construct a portfolio of exchange traded funds (ETFs) based on the financial statement data of their components. Although a number of ETFs and ETF-managed portfolios have emerged in the past few decades, the ability to apply neural networks to manage ETF portfolios is limited since the number and historical existence of ETFs are relatively smaller and shorter, respectively, than those of individual stocks. Therefore, we use the data of individual stocks to train our neural networks to predict the future performance of individual stocks and use these predictions and the portfolio deposit file (PDF) to construct a portfolio of ETFs. Multiple experiments have been performed, and we have found that our proposed method outperforms the baselines. We believe that our approach can be more beneficial when managing recently listed ETFs, such as thematic ETFs, of which there is relatively limited historical data for training advanced machine learning methods.
\end{abstract}


\maketitle
\pagestyle{plain}

\section{Introduction}

With recent progress in artificial intelligence, many researchers have applied advanced machine learning methods to predict the future price of the stock market or find a profitable trading signal\cite{Atsalakis2009,Cavalcante2016}. 
The predictability of the stock market has been a controversial topic for over half a century 
\cite{Fama1970,Malkiel2003,Jensen1978,Schumaker2009}.
 As a result, many researchers have attempted to prove that their approach is capable of yielding consistent profits.
 These approaches can be roughly divided into three groups: technical analysis-based groups, fundamental analysis-based groups, and alternative data-based groups.

The first group includes most of the previously mentioned studies \cite{Hu2015}. In these studies, the concept of technical analysis is basically extended \cite{Park2007,lo2000} by proposing novel machine learning models that primarily take the historical closing price and/or volume as an input feature \cite{Persio2016,Deng2016,Krauss2017,index2020}. The most important reason why so many previous studies fall into this group is that the historical closing price and volume data of individual stocks are the most common and abundant data for training advanced machine learning methods. The studies in the second group make use of the financial statement data of individual stocks to predict the future performance of the stock \cite{Huang2019}. Financial statement data usually refer to the data from a company's business activities and include income statements, balance sheets, and cash flow statements. Even though financial statement data are commonly used to evaluate the current value of a company \cite{Abarbanell1997}, there have been few attempts to apply advanced machine learning methods to predict the future performance of stocks based on financial statement data. Attempts have been made in the last group of studies to find the source of alpha, the profitable signal, from nontraditional data sources, such as investor sentiment in social networking service messages \cite{Bollen2011,Huang2014}, keywords extracted from news articles \cite{Ding2015,Schumaker2009}, or frequency changes of the specific words in search engine queries \cite{Da2011,Preis2013}. 

\begin{figure*} [h]
  \includegraphics[width=1.0\textwidth]{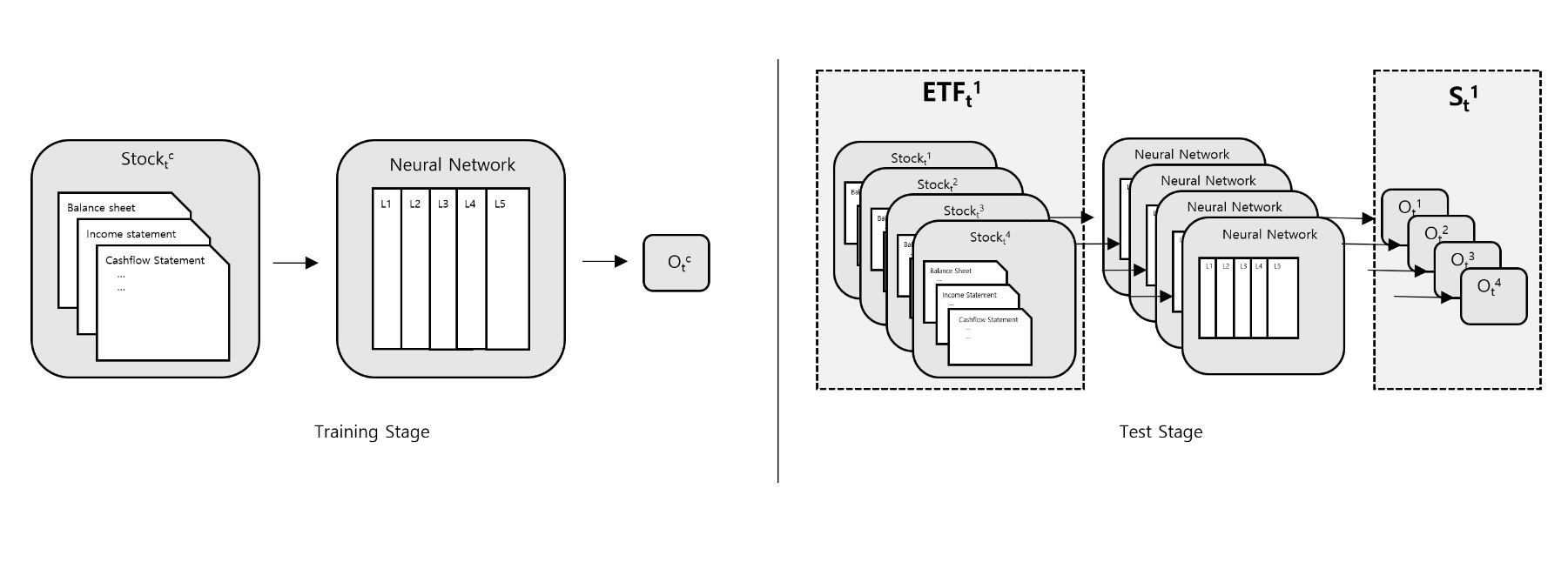}
  \caption{ Overall architecture of our method. }
  \Description{ Description }
  \label{fig:Fig1}
\end{figure*}

An ETF, like a regular stock, is an investment security that can be traded by investors in a stock exchange. An equity ETF consists of a group of stocks and operates much like mutual funds. Thus, investors who purchase ETFs can enjoy the benefit of diversifying their portfolio in a manner similar to those who have purchased a basket of individual stocks. This advantage as well as the advantage of low management fees has resulted in the ETF market explosion over the last two decades \cite{etf2020, Bourgi2018}. Despite the benefits and the presence of the ETF market, there are only a few studies \cite{Liew2018,Chen2006} that have attempted to construct portfolios of ETFs by applying advanced machine learning methods, such as deep neural networks. This is mainly because of the data shortage problem. Since the number of ETFs is relatively small compared with that of individual stocks and many ETFs have been recently listed, it is not easy to obtain a sufficient amount of data for training advanced machine learning methods.

In this study, we propose a novel method for constructing portfolios of ETFs based on the output of neural networks trained on the financial statement data of individual stocks. Our neural network is trained on listed stocks in the United States (US), which is a superset of components in our ETF investment universe. The output of our trained neural network and PDFs are used in the construction of portfolios of ETFs. The main contributions of this paper are as follows.

First, we mitigate the data shortage problem by using the data of the components of ETFs, not the data of ETFs, to train neural networks for constructing portfolios of ETFs. Otherwise, it is difficult to train and validate the neural networks using the data of ETFs, especially when the number of ETFs considered as an investment universe is small.
Moreover, it is impossible to apply an advanced machine learning method, which requires a certain number of historical data for training, to ETFs that have been recently listed.

In addition, we use financial statement data to construct portfolios of ETFs by making use of individual stock data. Financial statement data are considered a crucial factor in the evaluation of the current value of a stock or the prediction of its future performance. However, it is impossible to utilize financial statement data while constructing portfolios of ETFs unless the neural network is trained using individual stock data. To the best of our knowledge, no previous study used neural network trained on financial statement data for constructing a portfolio of ETFs.

\section{Method}
\subsection{Overview}
The key idea of our method is based on the fact that an ETF is simply an aggregation of its components. Therefore, the future performance of an ETF may be predictable if a machine learning model is capable of predicting the future performance of its components. The overall procedure of our method, which is illustrated in Fig \ref{fig:Fig1}, is designed to achieve this purpose and is described as follows.

The first step is to train our neural network to predict the future performance of individual stocks based on financial statement data. Our neural network takes the past 2 years, or 8 quarters, of financial statement data consisting of balance sheets, income statements, and cashflow statements of stock $\textit{c}$ as input and outputs $o_{t}^c$ at each time $\textit{t}$. Our neural network is trained to predict the 3-month ahead performance (the price change from $\textit{t}$ to $\textit{t+1}$) of the stock $\textit{c}$. This procedure is depicted at the left side of Fig \ref{fig:Fig1}. 

After training, scores are assigned to the individual stocks and the ETFs at the test stage. As depicted in the right side of Fig \ref{fig:Fig1}, 
our trained neural network assigns a score to each of the component of $ETF_{t}^1$ based on its financial statement data. Note that the four neural networks at the right side of Fig \ref{fig:Fig1}, are actually the same network with the same parameters. The purpose of the illustration is to emphasize that our neural network does not take the aggregated input of multiple stocks at once, it takes the input of the individual stocks independently. 
 
After assigning scores to all components of $ETF_{t}^1$, the score of $ETF_{t}^1$, which is $S_{t}^1$, is calculated based on the scores of individual stocks. By using the values ($o_{t}^1$, $o_{t}^2$, $o_{t}^3$, and $o_{t}^4$) from our neural network and the weights provided in PDF of $ETF_{t}^1$, score $S_{t}^1$ can be calculated in a quantitative manner. In other words, score $S_{t}^1$ is calculated by averaging the multiple of the score and weight of the individual stocks composing $ETF_{t}^1$.

After the scores are calculated for all of the ETFs in the investment universe at time $\textit{t}$, the portfolio of ETFs at $\textit{t}$ is constructed based on the scores of ETFs.

\subsection{Individual Stocks Data Description}

First, approximately 1200 of the top liquid stocks (based on data from Jan. 2021) listed on the US stock market are considered as the data for our experiments. To eliminate noise, at each time step, the suspended stocks and the stocks with no trading volume are eliminated. More specifically, the stocks that are not traded for longer than 5 days over the 3-month timespan are eliminated. Thus, the number of valid stocks that are used in our training and experiment changes over time.

\begin{figure} [h]
  \includegraphics[width=0.45\textwidth]{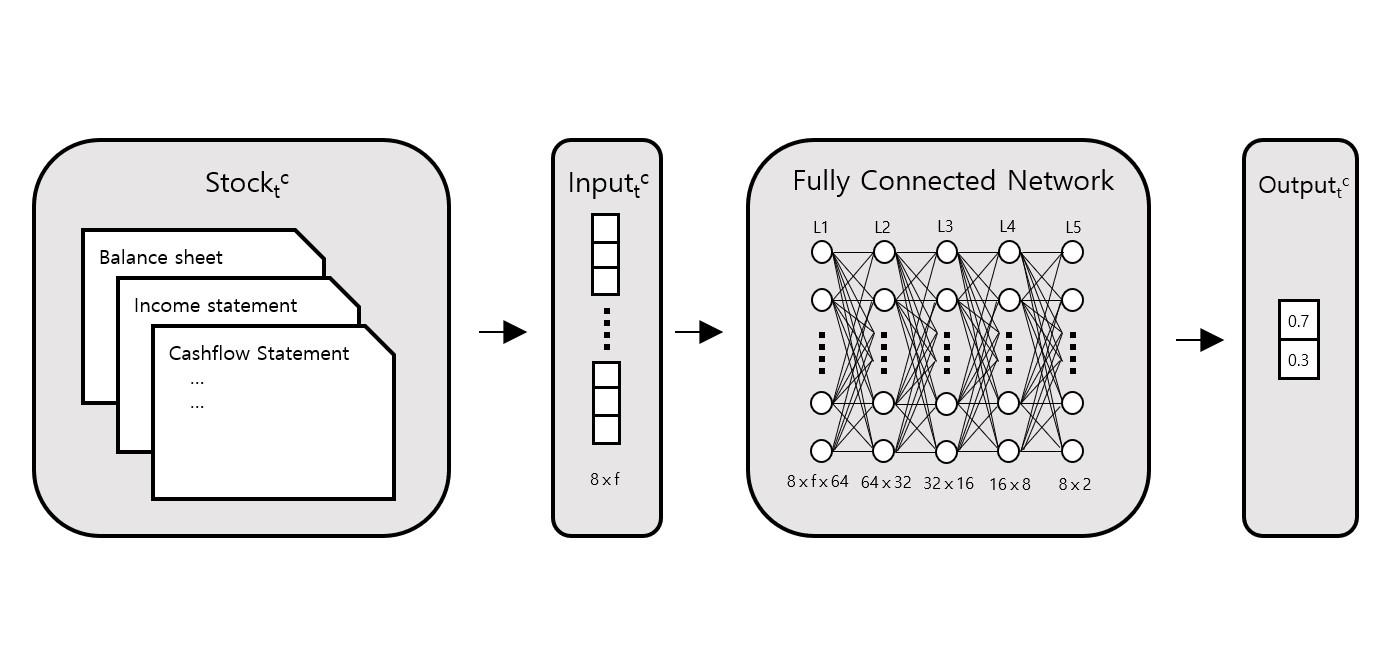}
  \caption{ Architecture of our neural network. }
  \Description{ Description }
  \label{fig:Fig2}
\end{figure}

After the stocks were chosen, the financial statement data\footnote{The type of financial statement data used in our method is listed in the appendix. } and daily closing price data were collected for these stocks. In the US, financial statement data are usually released quarterly. These releases occur approximately 2 to 12 weeks after the last business day of the previous quarter. In other words, financial statement data for the first quarter are generally released between the last day of March and June.
Therefore, to prevent using not released data as an input, the financial statement data for the first quarter are not used as an input feature until the last business day of June. 

Closing price data are used to generate labels on training stage, and evaluate the performance on test stage. Since our neural network is trained as a binary classification problem, the label is generated based on the 3-month ahead percent change of each stock. To avoid the long bias issue, the training data are neutralized by assigning "up" to the top 50\% of the companies and "down" to the bottom 50\% of the companies based on the 3-month ahead percent changes at each time $\textit{t}$. Considering the release period of financial statement data and operating issues, our neural network is trained on a quarterly basis. 

The data of individual stocks are collected over 22 years (Jan. 2000 - Apr. 2022) and divided into training and test sets. The first 10 years (Mar. 2002 - Mar. 2012) of data are used as training data, and the last 10 years (Mar. 2012 - Apr. 2022) of data are used as test data. Approximately 70\% of the training data are used for training our neural network, and the other 30\% of the data are used for optimizing the hyperparameters. All of the financial statement and closing price data are collected from Morningstar.

\subsection{ETF Data Description}
Similar to individual stocks, only the ETFs that are listed on the US stock market are considered as our investment universe. In addition, since individual stocks that are listed on the US stock market are used for training our neural network, ETFs that consist of individual stocks that are not listed in the US stock market are also excluded. 

Two groups of ETFs are selected for our investment universe\footnote{The description of our investment universe is provided in the appendix. }. First, 11 ETFs that track the 11 Global Industry Classification Standard (GICS) sectors in the US stock market are selected as the classic group. The ETF data in this group are collected over 10 years (Mar. 2012 - Apr. 2022), which is the same period as the test data from individual companies. Next, 26 ETFs with an inception date after 2017 are selected as the exotic group. The ETFs in the exotic group are selected based on their inception date and net asset value (NAV). The ETF data in the exotic group are collected starting from the inception date of each ETF to Apr. 2022.

\begin{figure} [h]
  \includegraphics[width=0.5\textwidth]{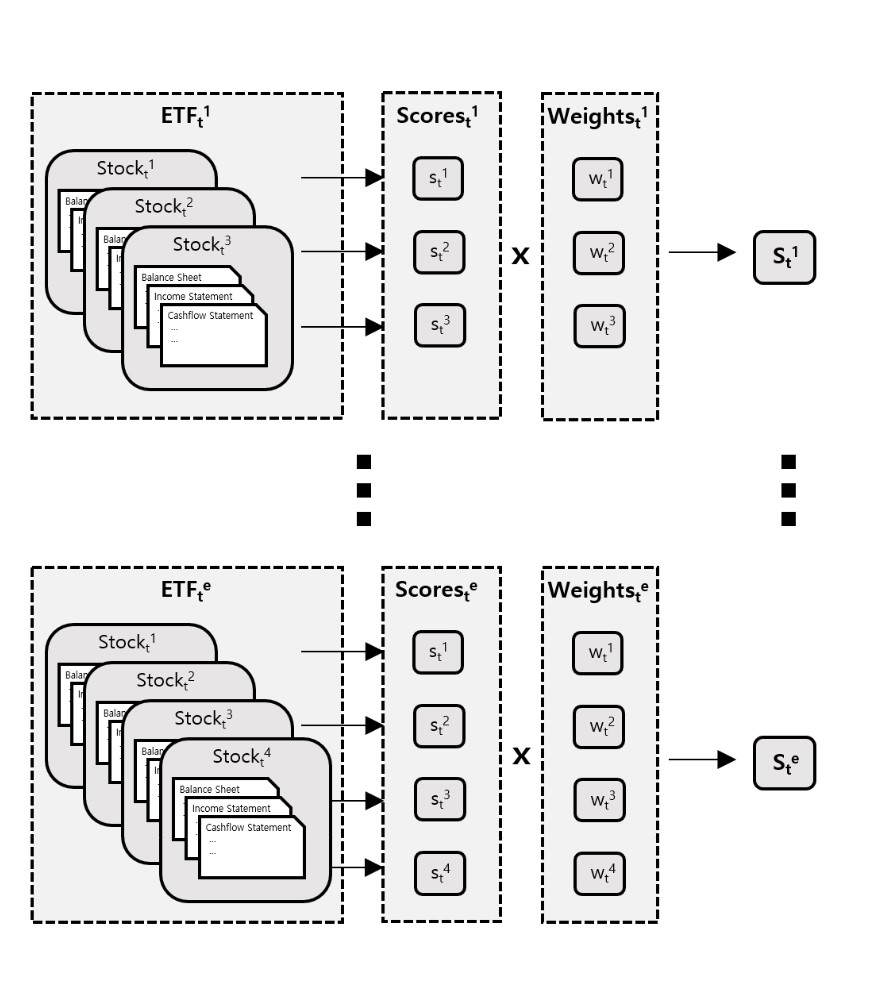}
  \caption{ Scoring ETFs. }
  \Description{ Description }
  \label{fig:Fig3}
\end{figure}

After the ETFs are chosen, the daily closing price data and PDF of each ETF are collected. The daily closing price is used to evaluate the performance of our method. The PDF provides the historical weight of each component (i.e., individual stocks) of the corresponding ETF. Entire PDFs and the closing price data of ETFs are also collected from Morningstar.

\subsection{Network architecture}
In this subsection, the shape of the input and the architecture of our neural network are provided. Among various choices of the network architecture, we attempted to keep the network architecture as simple as possible, since proposing a novel network architecture is not one of our main research contributions. The overall architecture of our neural networks is illustrated in Fig \ref{fig:Fig2}. 

The input of our neural network at time $\textit{t}$ is the past 8 quarters of financial statement data. The size of the input vector is $\textit{8}$ $\times$ $\textit{f}$, where $\textit{f}$ is the number of financial statement features.   

The architecture of our neural network consists of 5 fully connected layers followed by 1 softmax layer. Five fully connected layers consist of $\textit{8}$ $\times$ $\textit{f}$ $\times$ 64, 64 $\times$ 32, 32 $\times$ 16, 16 $\times$ 8 and 8 $\times$ 2 parameters. The first four fully connected layers contain a batch normalization \cite{Ioffe2015} layer followed by ReLU. Xavier initialization \cite{Glorot2010} is used for initializing the parameters in the network.

\subsection{Training using Individual Companies}
In the first stage of our method, our neural network is trained to predict the performance of the next quarter of individual stocks based on financial statement data.

\begin{algorithm}
\caption{ Training algorithm }
\begin{algorithmic}[1]

\FOR{ \textit{i} = \textit{0},\textit{maxiter} }
    \STATE Random sample batch of size $\textit{B}$ from training data. \\
    \STATE Calculate the loss \textit{L} in the batch. \\
    \STATE Perform gradient step to minimize  \textit{L} with respect to the parameter $\theta_{i}$. \\
    \IF{ \textit{i} > \textit{miniter} }
        \STATE At every \textit{C} iteration, save an evaluate the performance of the current parameter $\theta_{i}$. \\
    \ENDIF
\ENDFOR
\STATE Set $\theta_{*}$ as the best performing parameter.
\end{algorithmic}
\end{algorithm}

\subsubsection{Input Data Preprocessing}
Generally, the range of raw values of financial statement data to use as input of neural networks is extremely wide without preprocessing. There are multiple ways to preprocess input features before feeding them into neural networks. However, in our method, all the financial statement data are converted to percentage change to guide our neural networks to focus more on the changes in the feature values over time. The value of the $\textit{i}$th feature at time step $\textit{t}$ is calculated as follows.
 
 \begin{equation}
 \label{eq1}
  v_{t}^i= (r_{t}^i - r_{t-1}^i) / r_{t-1}^i
\end{equation}

\noindent where $v_{t}^i$ and $r_{t}^i$ refer to the input value and raw value of the $\textit{i}$th feature at time $\textit{t}$, respectively.

 \subsubsection{Loss function}
 
We define our problem as a binary classification problem and apply a standard supervised learning method to train our neural network. Our neural network is trained to minimize the negative log-likelihood loss. The loss of stock $\textit{c}$ at time $\textit{t}$ is calculated as follows.

 \begin{equation}
 \label{eq2}
 L_{t}^c = - \sum y_{t}^c \circ \log o_{t}^c
\end{equation}

\noindent where $y_{t}^c$ refers to an answer label, which is a one-hot vector of length 2, and $o_{t}^c$ refers to a softmax output of our neural network, which is also a vector of length 2. Each element of both vectors represents "up" and "down". The symbol $\circ$ represents elementwise multiplication. Thus, the loss of stock $\textit{c}$ at time $\textit{t}$ is summation over elementwise multiplication of $y_{t}^c$ and $o_{t}^c$.

\subsubsection{Training Process}

As mentioned in the Data Description section, the entire dataset of individual stocks is divided into training (Mar. 2002 - Mar. 2012) and test (Mar. 2012 - Apr. 2022) data. The training data are further divided into training (Mar. 2002 - Mar. 2009) and validation (Mar. 2009 - Mar. 2012) data. The hyperparameters\footnote{The values of the hyperparameters are listed in the appendix} are optimized by repeating the training and evaluation processes on the training and validation data. To prevent overfitting, the best performing parameter $\theta_{*}$ of our neural network is obtained based on the evaluation performance on the validation data. The pseudocode of the training process is described in Algorithm 1. The Adam optimizer \cite{Kingma2014} is used for taking the gradient step on loss \textit{L}.

\subsection{Scoring ETFs}

When the training process using individual stocks is complete, scoring ETFs using this trained neural network is quite straightforward. Fig \ref{fig:Fig3} provides an illustration of the overall process for scoring ETFs, which is described as follows.

First, for each $ETF_{t}^e$, our trained neural network assigns a score to the components of $ETF_{t}^e$, as follows.

 \begin{equation}
 \label{eq3}
 s_{t}^c = o_{t}^c[up]
\end{equation}

\noindent where $s_{t}^c$ refers to the score of stock $\textit{c}$ at time $\textit{t}$ and $o_{t}^c[up]$ refers to the softmax output value for the label "up" of stock $\textit{c}$ at time $\textit{t}$. In other words, the probability of stock $\textit{c}$ being classified as "up" at time $\textit{t+1}$ is assigned as a score of the stock.

Next, the score of $ETF_{t}^e$ is calculated based on the score and weight of its components.

 \begin{equation}
 \label{eq4}
 S_{t}^e = \sum_{c=1}^{m^e} w_{t}^c \times s_{t}^c
\end{equation}

\noindent where $w_{t}^c$ and $m^e$ refer to the weight of stock $\textit{c}$ and the number of components in $ETF_{t}^e$, respectively. The value $w_{t}^c$ is obtained from the PDF.

\begin{table}
  \caption{Experimental results on individual stocks from Mar. 2012 to Apr. 2022.}
  \label{tab:Tab1}
  \begin{tabular}{ccc}
    \toprule
    Portfolio & Annual Return (\%) \\
    \midrule
    S\&P 500 & 11.72 \\
    EW & 11.74  \\
    Top 80\% & 12.49 \\
    Top 60\% & 12.67 \\
    Top 40\% & 13.25 \\
    Top 20\% & 14.32 \\
  \bottomrule
\end{tabular}
\end{table}

\section{Experimental Results}
\subsection{Experimental Settings}
To evaluate the performance of our method, we conduct an experiment on our test period with two groups of ETFs. Since the performance of our trained neural network on individual stocks is crucial for the performance on ETFs, we also conduct an experiment on individual stocks. The test period is from Mar. 2012 to Apr. 2022 for experiments on both individual stocks and the classic ETF group and from June 2020 to Apr. 2022 for the exotic ETF group. We select the starting date of the experiment on the exotic ETF group based on the number of valid ETFs\footnote{See appendix for the list of inception dates. }. From June 2020, the number of ETFs in the exotic group exceeded 10.

In our experiments, we assume that the rebalancing interval is 3 months and conduct rebalancing on the last business days of March, June, September, and December. As described in the Individual Stocks Data Description section, only accessible financial statement data at each rebalancing time are used as input for our trained neural network. There are two main reasons why we fixed our rebalancing interval as 3 months. The first reason is that our neural network is trained based on financial statement data of individual stocks, and the release period of the financial statement data is 3 months. Another reason is that an excessively high turnover ratio may result in high trading costs and difficulty in operation, especially when managing large amount of  asset.

For our experiments, a portfolio is constructed based on the score described in the previous section as follows. First, the ETFs (or individual stocks) in the investment universe for each experiment are sorted based on their scores at each time \textit{t}. Here, only valid ETFs or individual stocks are considered. Next, top scoring ETFs (or individual stocks) are selected and invested for the next 3 months.

\subsection{Experiments on Individual Stocks}
As mentioned in the Methods section, our method of scoring ETFs is based on the softmax output of our trained neural network. Therefore, before conducting experiments on ETFs, validating whether the stocks with higher softmax output values outperform the stocks with lower softmax output values is necessary.

\begin{table}
  \caption{Experimental results on the classic ETF group from Mar. 2012 to Apr. 2022.}
  \label{tab:Tab2}
  \begin{tabular}{cccc}
    \toprule
    Portfolio & Annual Return (\%) & Volatility & Sharpe Ratio\\
    \midrule
    S\&P 500 & 11.72 & 16.41 & 0.71 \\
    EW & 9.70 & 15.87 & 0.61 \\
    Top 8 & 10.23 & 16.02 & 0.64 \\
    Top 7 & 10.36 & 16.08 & 0.64\\
    Top 6 & 11.41 & 15.94 & 0.72\\
    Top 5 & 12.34 & 15.93 & 0.78\\
    Top 4 & 12.51 & 16.27 & 0.77\\
    
  \bottomrule
\end{tabular}
\end{table}

 The portfolio is constructed as follows. At each rebalancing time $\textit{t}$, all the valid stocks at time $\textit{t}$ are sorted based on their score $s_{t}^c$, assigned by the output of our trained neural network. Then, the top $\textit{K}$ \% of the valid stocks are selected and invested until time $\textit{t+1}$, which is the next rebalancing time. An equal weight is assigned to each of the selected stocks. Our experimental period is approximately 10 years (Mar. 2012 - Apr. 2022), and the average number of valid stocks is 1017. Note that the number of valid stocks changes over time.   

 Table \ref{tab:Tab1} provides a summary of the experimental results on individual stocks. Each row represents a portfolio. For example, Top 80$\%$ refers to the portfolio consisting of the top 80\% of all valid stocks with the highest score. We compare our annualized return with two baselines, the S\&P 500 index and equal weight portfolio, which is exactly the same as the Top 100$\%$ portfolio.

The results clearly show that when the portfolio assigns more of its asset to the companies with higher scores, the annual return is increased.

\subsection{Experiments on ETFs}

In this section, the experimental results on two different groups of ETFs are provided. As mentioned in the ETF Data Description section, these two groups are the classic group and the exotic group. The purpose of conducting experiments on two different groups is as follows. 

 First, the ETFs in the classic group have a relatively long history and track 11 GICS sectors in the US stock market. Additionally, the components of these 11 ETFs are mutually exclusive and cover a large portion of the major stocks in the US stock market. Therefore, we reason that this group is more appropriate for evaluating the general performance of our method for a long period of time. In contrast, the ETFs in the exotic group usually consist of more thematic stocks, and the length of the historical data for these stocks is short. Thus, naively applying advanced machine learning methods, such as neural networks, to this group of ETFs is literally impossible. 

 By conducting experiments on two different types of ETFs, we attempted to show that our method generally outperforms the baselines over a long period of time, and unlike most of the previous works that use the same types of data on training and test stages, our method can be deployed for constructing portfolios of ETFs that have an extremely short period of historical time series for training neural networks. For this reason, we have not compared our method with previous works. Since most of the existing studies require a certain amount of data for training the proposed machine learning model, applying those studies to our dataset is not feasible.

\begin{table}
  \caption{Annual experimental results on the classic ETF group.}
  \label{tab:Tab3}
  \begin{tabular}{ccccccc}
    \toprule
    
    Year & \multicolumn{2}{c}{S\&P 500} & \multicolumn{2}{c}{EW} & \multicolumn{2}{c}{Top 4} \\
    \midrule
     & Return & Volatility & Return & Volatility & Return & Volatility \\
    \midrule
    2012 & 5.72 & 13.53 & 6.59 & 12.87 & \textbf{8.17} & 14.86 \\
    2013 & 23.13 & 10.63 & 19.46 & 10.78 & 19.01 & 10.73 \\
    2014 & 12.29 & 11.11 & 11.56 & 10.53 &\textbf{14.09} & 10.03 \\
    2015 & -1.03 & 15.27 & -4.32 & 14.73 & \textbf{0.23} &  14.61 \\
    2016 & 12.34 & 12.78 & 14.01 & 13.02 & 10.79 &  12.09 \\
    2017 & 17.96 & 6.56 & 12.05 & 6.61 & \textbf{13.35} &  6.90 \\
    2018 & -5.69 & 16.67 & -8.24 & 14.81 & -9.39 &  16.04 \\
    2019 & 26.87 & 12.25 & 21.65 & 10.96 & \textbf{23.31} &  11.52 \\
    2020 & 18.74 & 33.79 & 12.81 & 34.16  & \textbf{22.57} &  34.03\\
    2021 & 26.73 & 12.77 & 24.65 & 12.14 & \textbf{33.23} &  14.86 \\
    2022 & -13.48 & 21.84 & -7.52 & 17.21 & \textbf{-3.07} &  17.58 \\

  \bottomrule
\end{tabular}
\end{table}

The construction of portfolios of ETFs is very similar to that for portfolios of individual stocks. First, at each rebalancing time $\textit{t}$, all valid ETFs are sorted based on their score $S_{t}^c$, and the top $\textit{K}$ \% of valid ETFs or top $\textit{K}$ valid ETFs are selected and invested for the next 3 months. The equal weight is assigned to the selected ETFs.

\subsubsection{ The classic group}

Table \ref{tab:Tab2} provides a summary of the results of our experiment on the classic group. Note that $\textit{K}$ in this experiment is a number, not a percentage, because the inception dates of all of the ETFs in the class group are prior to the start date of test period, so the number of valid ETFs is always 11 for the entire test period. Thus, for example, the top 4 portfolio indicates that the portfolio consists of four ETFs with the highest score at each rebalancing time $\textit{t}$.

Similar to the experiment on the individual stocks, Table \ref{tab:Tab2} shows that when the portfolio consists of the ETFs with higher scores, the annual return is increased. All the portfolios consisting of selected ETFs outperform the equal weighted portfolio (portfolio EW), and two portfolios, top 5 and top 4, also outperform the S\&P 500 index in terms of annual return. It is worth noting that the annual return of the S\&P 500 index and equal weighted portfolio is quite different. This implies that the heavily weighted sectors in the S\&P 500 index have outperformed the other sectors for the last decade. In addition, Table \ref{tab:Tab2} shows another trend: the volatility of the portfolio is not affected much by the number of selected ETFs in the portfolio. Therefore, the Sharpe ratio also increases as the number of selected ETFs in the portfolio decreases. For simplicity, we assume that the risk-free rate is zero when calculating the Sharpe ratio.

The annual breakdown of the experimental results of the S\&P 500 index , equal weighted portfolio, and top 4 portfolio is described in Table \ref{tab:Tab3}. The result of the first and the last row, 2012 and 2022, is the result from Mar. 2012 to Dec. 2012 and from Jan. 2022 to Apr. 2022, respectively. As shown in Table \ref{tab:Tab3}, except for 2013, 2016, and 2018, the top 4 portfolio outperforms the equal weight portfolio. This result implies that the performance of our method is not biased in the short period of time. 

\subsubsection{The exotic group}

Table \ref{tab:Tab4} provides a summary of the results of our experiment on the exotic group. Note that the exotic group consists of 26 recently listed ETFs. Since the inception dates of these ETFs are mostly after 2020, the experiment on the exotic group is conducted on the data from Jun. 2020 to Apr. 2022. 

A similar tendency can be observed from Table \ref{tab:Tab4}. All the top $\textit{K}$ \% portfolios outperform the equal weight portfolio, and the top 40 \% and top 20 \% portfolios outperform the S\&P 500 index. The annual return is increased as the number of selected ETFs in the portfolio is decreased. However, volatility remains.

\begin{table}
  \caption{Experimental results on the exotic group.}
  \label{tab:Tab4}
  \begin{tabular}{cccc}
    \toprule
    Portfolio & Annual Return (\%) & Volatility & Sharpe Ratio\\
    \midrule
    S\&P 500 & 16.47 & 15.78 & 1.04 \\
    EW & 15.81 & 16.08 & 0.98 \\
    Top 80\% & 16.05 & 16.03 & 1.00 \\
    Top 60\% & 16.37 & 16.25 & 1.01\\
    Top 40\% & 16.64 & 16.51 & 1.01\\
    Top 20\% & 18.30 & 16.33 & 1.12\\
  \bottomrule
\end{tabular}
\end{table}

\section{Conclusion}

In this work, we propose a method for constructing ETF portfolio using neural networks trained on financial statement data of individual stocks. To the best of our knowledge, this work is the first to use the financial statement data to train neural networks for constructing ETF portfolio. Our method is conceptually simple but can effectively address the data shortage problem, which is crucial when applying an advanced machine learning method to construction the portfolio of ETFs. We conduct our experiment on two different groups of ETFs to support our contribution and show that our method generally outperforms the baselines. Based on our method, various types of machine learning methods could be applied for constructing ETF portfolios.

\bibliographystyle{ACM-Reference-Format}
\bibliography{reference_list}

\appendix

\begin{table}[!b]
  \caption{List of financial statement data.}
\begin{tabular}{l}
    \toprule
     Name  \\
    \midrule
    Total Revenue \\
    Operating Income \\
    Net Income \\
    Total Asset \\
    Current Asset \\
    Total Equity \\
    Current Liabilities \\
    Invested Capital \\
    Free Cashflow \\
    Operating Cashflow \\
    Market Capital \\
    
  \bottomrule
\end{tabular}
\end{table}

\begin{table}[!b]
  \caption{List of hyperparameters.}
  \begin{tabular}{lll}
    \toprule
    Name & Description & Value \\
    \midrule
    maxiter & Number of training iterations & 100,000 \\
    miniter & Number of iterations to start evaluation & 50,000 \\
    \textit{B} & Batch size & 128 \\
    \textit{C} & Saving interval & 1,000 \\
    \textit{learning rate} & The learning rate for Adam optimizer & 0.00025 \\
  \bottomrule
\end{tabular}
\end{table}

\begin{table}[!b]
  \caption{List of ETFs in the classic group. The column NAV is based on the data from June 2022.}
  \begin{tabular}{llll}
    \toprule
    Ticker & ISIN  & Inception Date & NAV (Million \$)\\
    \midrule
    IYW&	US4642877215&	2000-05-19&	6672.302\\
    IYZ&	US4642877132&	2000-05-26&	381.0265\\
    IYF&	US4642877884&	2000-05-31&	1911.769\\
    IYE&	US4642877967&	2000-06-16&	2685.947\\
    IYH&	US4642877629&	2000-06-16&	2875.589\\
    IYK&	US4642878122&	2000-06-16&	1326.637\\
    IYM&	US4642878387&	2000-06-20&	998.3269\\
    IDU&	US4642876977&	2000-06-20&	982.0283\\
    IYC&	US4642875805&	2000-06-28&	728.4399\\
    IYJ&	US4642877546&	2000-07-14&	1109.156\\
    IYR&	US4642877397&	2000-06-19&	4424.797\\
  \bottomrule
\end{tabular}
\end{table}

\begin{table}[!b]
  \caption{List of ETFs in the exotic group. The column NAV is based on the data from June 2022.}
  \begin{tabular}{llll}
    \toprule
     Ticker & ISIN & Inception Date & NAV (Million \$)\\
    \midrule
    ESGA&	US0250727528&	2020-07-15&	123.562\\
    CACG&	US5246821012&	2017-05-03&	158.704\\
    FLQS&	US35473P8766&	2017-04-26&	14.542\\
    FLV&	US0250727940&	2020-04-02&	215.247\\
    FMIL&	US3160923609&	2020-06-04&	53.786\\
    WWOW&	US25460G6411&	2020-12-17&	4.064\\
    ESMV&	US46436E4456&	2021-11-04&	4.457\\
    XVV&	US46436E5693&	2020-09-24&	215.225\\
    EQOP&	US63875W1099&	2020-09-17&	8.335\\
    ESGV&	US9219107334&	2018-09-20&	5672.33\\
    AVDG&	US64157X2036&	2020-12-30&	9.33\\
    WLTG&	US26923N8011&	2021-12-07&	9.417\\
    GK&	    US00768Y3707&	2021-07-02&	17.009\\
    GPAL&	US38149W7478&	2021-12-15&	6.487\\
    UGCE&	US90431R2094&	2021-04-16&	28.78\\
    AVDR&	US64157X1046&	2020-12-30&	23.621\\
    FBCV&	US3160923450&	2020-06-04&	106.376\\
    XJH&	US46436E5511&	2020-09-24&	71.098\\
    MID&	US0250727601&	2020-07-15&	22.606\\
    BOUT&	US45782C7636&	2018-09-12&	15.294\\
    QGRO&	US0250723071&	2018-09-12&	206.319\\
    FLQM&	US35473P8840&	2017-04-26&	57.873\\
    USXF&	US46436E7673&	2020-06-18&	566.202\\
    XJR&	US46436E5446&	2020-09-24&	28.288\\
    STLV&	US46436E3045&	2020-01-16&	6.617\\
    STLG&	US46436E4035&	2020-01-16&	1.517\\
  \bottomrule
\end{tabular}
\end{table}

\end{document}